\def\ket#1{\left|#1\right\rangle}

\documentclass[aps,prl,twocolumn,floatfix,10pt]{revtex4-1} 
\usepackage{graphicx,amsmath,amssymb,ulem}
\usepackage{color}

\begin{document}

\title{Spontaneous breaking of spatial and spin symmetry in spinor condensates}

\author{M.~Scherer$^1$, B.~L\"ucke$^1$, G.~Gebreyesus$^2$, O.~Topic$^1$, F.~Deuretzbacher$^2$, W.~Ertmer$^1$, L.~Santos$^2$, J.J.~Arlt$^3$, C.~Klempt$^1$}

\affiliation{$^1$ Institut f\"ur Quantenoptik, Leibniz Universit\"at Hannover, 30167~Hannover, Germany}
\affiliation{$^2$ Institut f\"ur Theoretische Physik, Leibniz Universit\"at Hannover, 30167~Hannover, Germany}
\affiliation{$^3$ QUANTOP, Department of Physics and Astronomy, University of Aarhus, 8000 \AA{}rhus C, Denmark}

\date{\today}

\begin{abstract}
Parametric amplification of quantum fluctuations constitutes a fundamental mechanism for spontaneous symmetry breaking. In our experiments, a spinor condensate acts as a parametric amplifier of spin modes, resulting in a twofold spontaneous breaking of spatial and spin symmetry in the amplified clouds. Our experiments permit a precise analysis of the amplification in specific spatial Bessel-like modes, allowing for the detailed understanding of the double symmetry breaking. On resonances that create vortex-antivortex superpositions, we show that the cylindrical spatial symmetry is spontaneously broken, but phase squeezing prevents spin-symmetry breaking. If, however, non-degenerate spin modes contribute to the amplification, quantum interferences lead to spin-dependent density profiles and hence spontaneously-formed patterns in the longitudinal magnetization.

\end{abstract}

\maketitle

Spontaneous symmetry breaking is fundamental in disparate scenarios in physics ranging from cosmology~\cite{Kibble1976} and  
particle physics~\cite{Turok1990} to liquid crystals~\cite{Chuang1991} and superfluid Helium~\cite{Zurek1985}. 
Symmetry breaking is also crucial in Bose-Einstein condensates (BECs), 
where the $U(1)$ symmetry is spontaneously broken. Even more interesting, quantum gases provide 
unprecedented possibilities for the investigation of non-equilibrium dynamics, and in particular for the detailed analysis 
of dynamical symmetry breaking, including the formation of topological defects via the Kibble-Zurek 
mechanism~\cite{Kibble1976,Zurek1985,Sadler2006}.

Spinor gases, formed by atoms with non-zero spin, present an exceptionally rich physics due to the non-trivial interplay between internal and external degrees of freedom~\cite{Ho1998,Ohmi1998,Schmaljohann2004,Chang2004}. A remarkable scenario is provided by BECs initially prepared in an unstable $m_F=0$~($\ket{0}$) Zeeman state. Spin changing collisions lead to the correlated creation of atom pairs in $m_F=\pm 1$~($\ket{\pm 1}$) in a process resembling optical down conversion in nonlinear crystals. As a result, parametric amplification of matter waves in spinor BECs has been recently observed~\cite{Leslie2009,Klempt2010}. 

Parametric amplification of quantum fluctuations constitutes a fundamental mechanism for symmetry breaking, and hence spinor BECs constitute an exciting system to investigate spontaneously broken symmetries. In a recent seminal experiment~\cite{Sadler2006} the sudden quench of a spinor BEC from a polar into a ferromagnetic phase was followed by the formation of ferromagnetic domains and topological defects in the transverse magnetization, whereas the longitudinal magnetization remained negligible. Ref.~\cite{Sadler2006} provided a major insight in the formation of topological defects, but the nature of the symmetry breaking mechanism remained largely unexplored. 

This article explores the nature of the symmetry breaking in a $^{87}$Rb $F=2$ BEC initially prepared in the $\ket{0}$ state. In our experimental system the $\ket{\pm 1}$ atoms are amplified on well resolved magnetic field resonances~\cite{Klempt2009} with characteristic Bessel-like spatial modes, since the energy splitting of these modes is on the same scale as the spin changing interactions. Our experiments hence permit a precise analysis of the amplification of quantum spin fluctuations, allowing for a detailed characterization of the symmetry breaking mechanism. Interestingly, a twofold spontaneous breaking of spatial and spin symmetries may occur. On one hand, we show that quantum fluctuations of the relative phase between amplified degenerate vortex/antivortex states may break the cylindrical symmetry imposed by the trap. On the other hand, contrary to the situation in Ref.~\cite{Sadler2006}, the density profiles in $\ket{\pm 1}$ may differ from each other, leading to spontaneously-formed longitudinal magnetization patterns only if various non-degenerate spin modes are significantly amplified. We show that this novel type of spin-symmetry breaking is linked to quantum interferences occurring during the amplification.

In the following we consider the initial BEC in $\ket{0}$ as a classical field $\sqrt{n_0(\vec r)}$, whereas the excitations in $\ket{\pm 1}$ are represented by the operators $\delta\hat\psi_{\pm 1}(\vec r)$~\cite{Klempt2009}. If the population in $\ket{\pm 1}$ remains small compared to that in $\ket{0}$, the dynamics may be described by the linear Hamiltonian
\begin{equation}
 \hat H\!=\!\!\!\int d^3{r}\!\!\!\sum_{m=\pm 1}\!\!\delta\hat\psi_{m}^\dag \hat H_{eff}\delta\hat\psi_{m}
\!+\Omega_{eff}\! \left [
\delta\hat\psi_{1}^\dag \delta\hat\psi_{-1}^\dag+h.c. \right ]
\label{eq:H}
\end{equation}
where $\hat H_{eff}=-\hbar^2\nabla^2/2M+V_{eff}(\vec r)+q$, $M$ is the atomic mass, and $\Omega_{eff}(\vec r)=U_{1} n_0(\vec r)$ characterizes the pair creation due to spin changing collisions. The effective potential experienced by $\ket{\pm 1}$ atoms is $V_{eff}(\vec r)\equiv V(\vec r)+(U_{0}+U_{1})n_0(\vec r)-\mu$, with the harmonic trap $V(\vec r)$, and the chemical potential $\mu$ of the initial BEC. 
The interaction strengths are given by $U_0$ and $U_1$~\cite{Ho1998,Klempt2010}. The magnetic field $B$ influences the dynamics only due to the quadratic Zeeman energy, $q \propto B^2 < 0$.

In our experiments the initial BECs are produced in a cylindrically symmetric trap well within the Thomas-Fermi regime. Hence $V(\vec r)+U_{0}n_0(\vec r)-\mu\simeq 0$ holds within the BEC region $r<r_{tf}$, where $r_{tf}$ is the Thomas-Fermi radius. Therefore $V_{eff}$ is flat, apart from the small repulsion $U_{1}n_0(\vec r)$ which produces a bump of $\sim h \times 30$~Hz in the center. Beyond the BEC surface, $V_{eff}(\vec r)=V(\vec r)$ equals the external potential and rises sharply. Hence the radial potential may well be approximated by a 2D circular box potential with radius $\simeq r_{\rm tf}$. In our experiments the energy splitting between transverse 
modes is larger than the typical energy of spin-changing collisions, and hence the transverse modes are well defined. On the contrary, 
the energy spacing of the accessible axial modes is much smaller than the spin-changing interaction, and hence they are not resolved. 
Our detection integrates along the axial direction, and hence a 2D description based on the transversal modes is sufficient. 
The single-atom eigenfunctions of the 2D circular box are
\begin{equation}
\varphi_{nl}(r,\phi)=\frac{1}{\sqrt{\pi} r_{\rm tf} J_{l+1}(\beta_{nl})}J_l \left ( \beta_{nl} \frac{r}{r_{\rm tf}} \right ) e^{i l \phi}
\label{eq:eigenfunctions}
\end{equation}
with eigenenergies $\epsilon_{nl}=\hbar^2 \beta_{nl}^2/2 m r_{\rm tf}^2$. Here, $J_l$ are Bessel functions of the first kind and $\beta_{nl}$ is the $n$th zero of $J_l$. Each mode is identified by two quantum numbers, $n$ and $l$, for the radial excitation and the angular momentum.

A deeper understanding is gained by expanding~(\ref{eq:H}) in these eigenfunctions: $\delta\hat\psi_m(\vec r)=\sum_{nl} \varphi_{nl}(\vec r) \hat a_{n,l,m}$. The Hamiltonian then splits into 
$\hat H=\sum_{n,|l|} H_{n|l|}$, where 
\begin{eqnarray}
\hat H_{n|l|}&=&(\epsilon_{n|l|}+q)\sum_{m,l=\pm|l|} \hat{a}_{n,l,m}^\dag \hat a_{n,l,m} \nonumber \\ 
&+&\Omega \left(\hat a_{n,l,1}^\dag\hat a_{n,-l,-1}^\dag + \hat a_{n,-l,1}^\dag\hat a_{n,l,-1}^\dag + {\rm h.c.}\right ), 
\label{eq:Hnl}
\end{eqnarray}
with $\Omega=U_1 n_0$~\footnote{For simplicity we use a constant $\Omega$ within the BEC region. 
The spatial dependence of $\Omega$ leads to a coupling between modes, which does not change the qualitative picture.}. The mode is stable if the eigenenergy $\xi_{nl}=\sqrt{(\epsilon_{nl}+q)^2-\Omega^2}$ of $\hat H_{nl}$ is real, whereas an imaginary $\xi_{nl}$ leads to an exponentially increasing population of $\varphi_{nl}$. 
Resonances in the spin dynamics occur when $\epsilon_{nl}+q=0$. 
Since $\epsilon_{nl}$ depend on $|l|$, modes with the same $n$ and opposite $l$ and $-l$ 
are degenerate.

The experimental apparatus to observe these Bessel modes and to study the symmetry breaking was described previously~\cite{Klempt2008, Klempt2009}. Briefly, a BEC of $^{87}$Rb with a thermal fraction below $20\%$ is prepared in the $F = 2, m_F=0$ state in an optical dipole trap. Two intersecting laser beams at a wavelength of $1064$~nm provide a cylindrical trapping potential with measured oscillation frequencies of $187,183$ and $67$~Hz. Before starting the experiment a strong magnetic field gradient is used to expel residual atoms in $\ket{m\neq 0}$ states from the trap. This reduces unwanted atoms in $\ket{\pm 1}$ which may dominate the amplification~\cite{Klempt2010}. Subsequently, the magnetic field is set to a value between $0$ and $2.5$~G, and the spin components evolve for an adjustable hold time. 

%%%% FIGURE 1
\begin{figure}%[ht]
\centering
\includegraphics*[width=0.85\columnwidth]{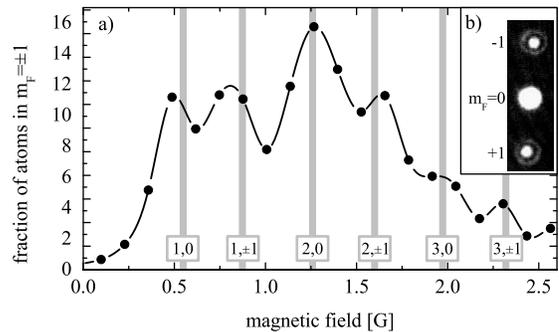}
\caption{(a) Fraction of atoms transferred into the $\ket{\pm 1}$ state within $18.5$~ms as a function of the applied magnetic field. Each data point is an average over $30$ realizations. The vertical gray lines indicate the resonance positions obtained from a 2D circular box model, and the labels indicate the corresponding Bessel modes. (b) Absorption image of a $\ket{0}$ BEC and the $\ket{\pm 1}$ clouds recorded at $1.29$~G.}
\label{fig1}
\end{figure}

After this spin evolution time, the trap is switched off to allow for ballistic expansion. During the first $2$~ms, all components expand together,
such that the $\ket{\pm 1}$ clouds experience the repulsion from the expanding $\ket{0}$ BEC. Once the densities are low enough, the spin components are separated by a magnetic field gradient applied for $3.5$~ms. After a final expansion time of $1.5$~ms, the 2D column densities are imaged along the weak trap axis as shown in Fig.~\ref{fig1} (b). 
Note that the expansion of the $\ket{\pm 1}$ clouds is governed by two contributions. The kinetic energy distribution in the trap and the repulsion from the $\ket{0}$ BEC during the initial expansion. Since the kinetic energy associated to the typical momentum $\sim 1/r_{\rm tf}$ is negligible compared to the mean-field repulsion, the clouds expand self-similarly~\cite{Castin1996,Kagan1996}. Therefore, the image obtained after time-of-flight expansion reproduces the density distribution in the trap.

%%%% FIGURE 2
\begin{figure}%[ht]
\centering
\includegraphics*[width=0.75\columnwidth]{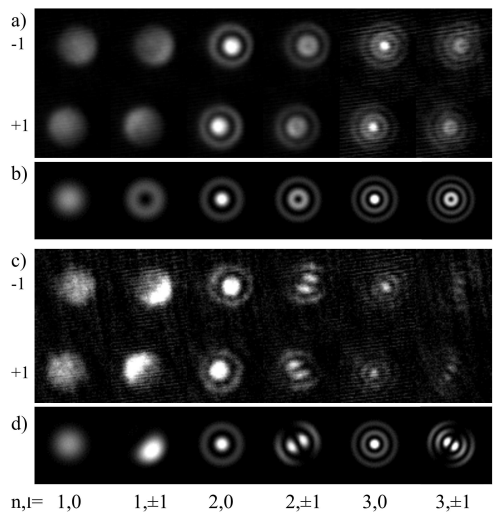}
\caption{Experimental and theoretical density distributions on the resonance positions 
after time-of-flight expansion. 
(a) Averaged experimental density profiles. 
(b) Calculated pure Bessel distributions corresponding to the experimental situation. 
(c) Individual experimental density profiles. 
(d) Calculated superpositions of Bessel distributions (see text). 
The $\ket{0}$ BEC was omitted in (a) and (c) for clarity.}
\label{fig2}
\end{figure}

Figure~\ref{fig1} shows the fraction of atoms in $\ket{\pm 1}$ as a function of the magnetic field. One observes six resonances, which arise when the quadratic Zeeman energy matches the eigenenergy of a 
mode $(n,l)$. These positions are indicated by gray lines, obtained from $\epsilon_{n,l}$ by fitting $r_{\rm tf}=3.9~\mu$m, in agreement with a calculated $3.7~\mu$m. This is reflected by the density distributions of 
the $\ket{\pm 1}$ clouds. Figure~\ref{fig2}~(a) shows the averaged density profiles of all realizations recorded at the six resonance positions, which match 
the theoretical Bessel distributions of Fig.~\ref{fig2}~(b) strikingly well. As expected the averaged profiles 
share the symmetries of the Hamiltonian~\eqref{eq:H}, i.e. a spatial cylindrical symmetry and a spin symmetry under the interchange $\ket{1}\leftrightarrow \ket{-1}$. Only the resonance for $(n,l)=(1,\pm 1)$, shows an orientation (see below).

However, the individual density profiles shown in Fig.~\ref{fig2}~(c) are clearly different. Since the $(1,0)$, $(2,0)$ and $(3,0)$ modes have $l=0$, the clouds remain cylindrically symmetric on the corresponding resonances even in individual realizations. The remaining resonances however correspond to $|l|=1$ modes, which are two-fold degenerate, allowing for a clockwise rotation (vortex) or an anti-clockwise one (anti-vortex). Since the $\ket{\pm 1}$ clouds result from amplified quantum fluctuations~\cite{Klempt2010}, the relative strength of vortex/antivortex modes is chosen randomly for each realization. If just one of the modes is populated, the result is a cylindrically symmetric density distribution. However, if superpositions of the two modes are populated, non cylindrically-symmetric standing radial waves emerge with a clear orientation (Fig.~\ref{fig2}~(c) $(2,\pm 1)$ and $(3,\pm 1)$). The angle of the orientation depends on the phase difference between the vortex and anti-vortex modes, which is randomly chosen by quantum fluctuations in individual realizations, constituting a clear example of spontaneous symmetry breaking. 

%%%% FIGURE 3
\begin{figure}%[ht]
\centering
\includegraphics*[width=0.75\columnwidth]{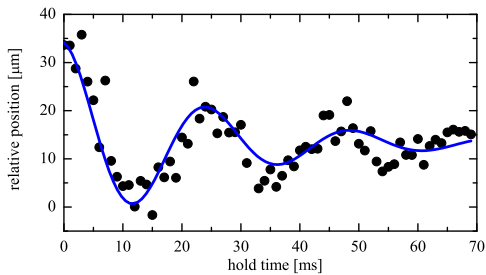}
\caption{Relative position of the $\ket{\pm 1}$ clouds prepared at $0.64$~G after variable hold time at high magnetic field. A fit with a damped oscillation yields a frequency of $40\pm3$~Hz, which closely corresponds to the energy difference between the resonances corresponding to the $(1,0)$ and $(1,\pm 1)$ modes.}
\label{fig3}
\end{figure}

This argument does not hold for density profiles recorded on the $(1,\pm 1)$ resonance. Since it is very close to $(1,0)$ mode the clouds are produced in superpositions of $(1,0)$ and $(1,\pm 1)$. As an example, a $50/50$ superposition is shown in Fig.~\ref{fig2}~(c) and~(d). Such superpositions are strongly off-centered and thus small magnetic field gradients are enough to break the cylindrical symmetry. To test this interpretation, $\ket{\pm 1}$ clouds were produced on this resonance and subsequently the magnetic field was set to a high value not supporting further spin dynamics. After a variable hold time, the relative center of mass positions of the $\ket{\pm 1}$ clouds was recorded. Figure~\ref{fig3} shows that the two clouds perform a damped oscillation at $40\pm3$~Hz, which matches the energy difference between the $l=0$ and $|l|=1$ resonances of $37\pm5$~Hz. This confirms the am\-pli\-fi\-ca\-tion of a coherent superposition including the $(1,0)$ mode.

Symmetry breaking is therefore best studied on the resonance corresponding to the $(2,\pm 1)$ mode. To evaluate 
the symmetry breaking process statistically, $2000$ images were taken at four magnetic field values between $1.58$~G and $1.84$~G for an evolution time of $21.5$~ms. For each picture, the angle of orientation of the $\ket{\pm 1}$ clouds was determined by two independent methods. First, a superposition of the two counter-rotating Bessel modes was fitted to the density distribution. Second, the eigenvectors of the quadrupole moment tensor of the density distribution were determined. The difference of the two angle measurements has a standard deviation of $15^\circ$ and $78\%$ of the measurements differing less than $40^\circ$ were included. The remaining $22\%$ were excluded from the analysis, since the total atom number was too small or the image was cylindrically symmetric or experimental noise corrupted one of the angle measurements. We checked that the experimental result remained stable even for a large variation of the exclusion conditions.

%%%% FIGURE 4
\begin{figure}%[ht]
\centering
\includegraphics*[width=0.88\columnwidth]{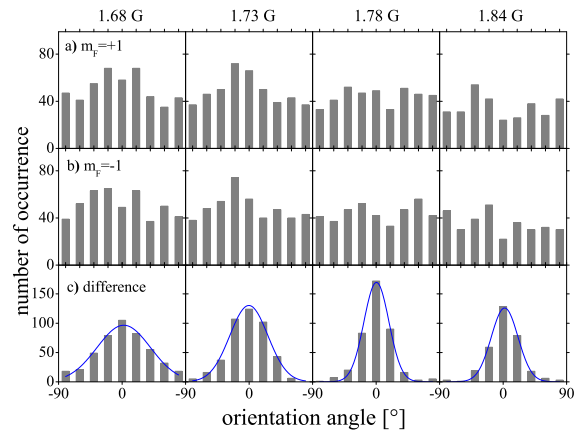}
\caption{Orientation of individual density distributions recorded in the vicinity of the $(2,1)$ mode. (a), (b) Distribution of angles for the $\ket{1}$ and $\ket{-1}$ states. (c) Distribution of the difference between the angles of orientation.}
\label{fig4}
\end{figure}

The results are presented in Fig.~\ref{fig4}. For each magnetic field, a histogram shows the distribution of angles for $\ket{1}$ and $\ket{-1}$ as well as the difference between these angles. All angles of the $\ket{\pm 1}$ clouds occur with roughly the same probability, showing that the cylindrical symmetry is indeed spontaneously broken. This confirms that neither the spurious production of $\ket{\pm 1}$ atoms~\cite{Klempt2010} nor slight asymmetries of the potential influence the population of higher modes and thus do not break the symmetry~\footnote{The slight accumulation of small angles at $1.73$~G is not significant but may indicate a slight asymmetry in the trap}. 

Interestingly, the relative angle (Fig.~\ref{fig4} (c)) shows an intriguing behavior. At $1.78$~G, the two angles are locked to each other within the angle measurement uncertainty. Towards lower or higher magnetic fields, the locking relaxes and differing angles become more probable. Such realizations with different orientations of the density distributions in $\ket{1}$ and $\ket{-1}$ break the spin symmetry of the density profiles. 
As a result, a spatial pattern of the longitudinal magnetization, 
$S_z=\sum_m m \delta\hat\psi_m^\dag(\vec r)\delta\psi_m(\vec r)$ emerges, contrary to Ref.~\cite{Sadler2006} where 
$S_z$ was negligible.

This novel spin-symmetry breaking cannot be explained only by the amplification of quantum fluctuations. 
The physical mechanism may be understood from Eq.~\eqref{eq:Hnl}, which for unstable modes may be re-written as
\begin{equation}
\hat H_{n,|l|}\! =\! \frac{|\xi_{n,|l|}|}{2}\!\!\!\!\sum_{l=\pm |l|,m} \!\!\left [\hat X_{n,l,m}^{(1)} \hat X_{n,l,m}^{(2)}\! + \! \hat X_{n,l,m}^{(2)} \hat X_{n,l,m}^{(1)} \right ],
\label{eq:HnU}
\end{equation}
where 
\begin{equation}
\hat X_{n,l,m}^{(1,2)}=\frac{\left ( e^{\pm i\theta_{n,l}}\hat a_{n,\pm l,\pm m}+e^{\mp i\theta_{n,l}} \hat a_{n,\mp l,\mp m}^\dag \right )}{(2\sin 2\theta_{n,l})^{1/2}},
\label{eq:Xnm12} \\
\end{equation}
with $\cos 2\theta_{n,l} = (\epsilon_{n,|l|}+q)/\Omega$. The operators $\hat X_{n,l,m}^{(1,2)}$ can be identified as the quadratures 
($[\hat X_{n,l,m}^{(1)},\hat X_{n,l,m}^{(2)}]=i$) corresponding to the amplification dynamics. The quadrature $\hat X_{n,l,m}^{(1)}$ is stretched ($\sim e^{|\xi_{n,|l|}|t}$), whereas $\hat X_{n,m}^{(2)}$ is squeezed ($\sim e^{-|\xi_{n,|l|}|t}$). Hence the amplification of the $(n,l)$-mode is characterized by 
the mixing angle $\theta_{n,l}$.

After a given time $t$ we may evaluate the density-density correlation functions $G_{m,m'}(\vec r)\equiv\langle n_m(\vec r)n_{m'}(\vec r)\rangle$
\begin{eqnarray}
G_{m,m}(\vec r)&=&2\sum_{\vec n,\vec n'} F_{\vec n,\vec n'}(\vec r), 
\label{eq:nm-nm}\\
G_{m,-m}(\vec r)&=&\sum_{\vec n,\vec n'} F_{\vec n,\vec n'}(\vec r,t) [1+\cos 2(\theta_{\vec n}-\theta_{\vec n'})], 
\label{eq:n1-n-1} 
\end{eqnarray}
where only unstable modes $\vec n\equiv (n,l)$ are summed, and 
\begin{equation}
F_{\vec n,\vec n'}(\vec r)=\frac{e^{2(|\xi_{\vec n}|+|\xi_{\vec n'}|)t}}{16\sin^2 2\theta_{\vec n} \sin^2 2\theta_{\vec n'}}
|\varphi_{\vec n}(\vec r)|^2 |\varphi_{\vec n'}(\vec r)|^2.
\label{eq:Fnn}
\end{equation}

Equation~\eqref{eq:n1-n-1} shows that spin symmetry breaking is caused by quantum interference between unstable modes with different mixing angles $\theta_{\vec n}$. This is the case at magnetic fields slightly above or below $1.78$~G, for which the $(2,0)$ or $(3,0)$ modes interfere with the strong $(2,\pm 1)$ modes. Note, however, that degenerate modes ($\epsilon_{n,l}=\epsilon_{n,-l}$) possess equal $\theta_{n,l}$, resulting in the observed locked angles of the $\ket{\pm 1}$ clouds exactly at $1.78$~G.

Alternatively, the angle locking may be explained by phase squeezing. At $1.78$~G only the $(2,\pm 1)$ modes contribute, and hence only $\hat H_{2,|1|}$ must be considered. At the resonance ($\epsilon_{2,|1|}+q=0$) the evolution operator $e^{-i\hat H_{2,|1|}t/\hbar}$ corresponds to a two-mode squeezing operator, which produces two output waves with a perfectly anti-squeezed phase difference but a dynamically squeezed phase sum~\cite{Barnett1990} when applied to the spin vacuum. For a sufficient evolution time, the phase sum approaches the squeezing angle~\cite{Barnett1990}, and hence the phases $\phi_{l=\pm 1, m}$ in the linear superposition of $(2,\pm 1)$ fulfill $\phi_{1,1}+\phi_{-1,-1}= \phi_{1,-1}+\phi_{-1,1}$. The orientation of the patterns is defined by the relative phase between the two counter-rotating modes in each $\ket{m}$ state, therefore the two angles are dynamically locked to each other since $\phi_{1,1}-\phi_{1,-1}=\phi_{-1,1}-\phi_{-1,-1}$. Again, the unlocking of the angles is due to the interference with neighboring resonances, which correspond to other Bessel modes. 

In summary, the amplification of $\ket{\pm 1}$ states is characterized by a non-trivial double spontaneous symmetry breaking. 
On one hand, the cylindrical spatial symmetry is broken by quantum fluctuations of the relative phase between vortex/antivortex modes, although 
phase squeezing may prevent the breaking of the spin symmetry. On the other hand, 
if various non-degenerate spin modes are amplified, spontaneously-formed longitudinal magnetization patterns appear due to quantum interferences between the different amplified quadratures.
Our results show that spinor gases constitute an exceptionally controllable system for the detailed analysis of symmetry 
breaking and in particular of its close connection to multimode squeezing during parametric amplification.

We acknowledge support from the Centre for Quantum Engineering and Space-Time Research QUEST, the European Science Foundation (EuroQUASAR) 
and the Danish National Research Foundation Center for Quantum Optics (QUANTOP).
\bibliography{Scherer}

\end{document}